\documentclass[%
 reprint,
superscriptaddress,
 amsmath,amssymb,
 aps,
 prb,
]{revtex4-2}
\usepackage{graphicx}% Include figure files
\usepackage{dcolumn}% Align table columns on decimal point
\usepackage{bm}% bold math
\usepackage{natbib}
\usepackage{hyperref,color}% add hypertext capabilities
\begin{document}

\preprint{APS/123-QED}

\title{Breaking a Bloch-wave interferometer: quasiparticle species-specific temperature-dependent nonequilibrium dephasing}% Force line breaks with \\

\author{Joseph B. Costello}
\affiliation{Physics Department, University of California, Santa Barbara, California 93106, USA}
\affiliation{Institute for Terahertz Science and Technology, University of California, Santa Barbara, California 93106, USA}
\author{Seamus D. O'Hara}%
\affiliation{Physics Department, University of California, Santa Barbara, California 93106, USA}
\affiliation{Institute for Terahertz Science and Technology, University of California, Santa Barbara, California 93106, USA}
\author{Qile Wu}
\affiliation{Physics Department, University of California, Santa Barbara, California 93106, USA}
\affiliation{Institute for Terahertz Science and Technology, University of California, Santa Barbara, California 93106, USA}
\author{Moonsuk Jang}
\affiliation{Physics Department, University of California, Santa Barbara, California 93106, USA}
\affiliation{Institute for Terahertz Science and Technology, University of California, Santa Barbara, California 93106, USA}
\author{Loren N. Pfeiffer}
\affiliation{Electrical Engineering Department, Princeton University, Princeton, New Jersey 08544, USA}
\author{Ken W. West}
\affiliation{Electrical Engineering Department, Princeton University, Princeton, New Jersey 08544, USA}
\author{Mark S. Sherwin}
\affiliation{Physics Department, University of California, Santa Barbara, California 93106, USA}
\affiliation{Institute for Terahertz Science and Technology, University of California, Santa Barbara, California 93106, USA}

\date{\today}% It is always \today, today,

\begin{abstract}
Recently, high-order sideband polarimetry has been established as an experimental method that links the polarization of sidebands to an interference of Bloch wavefunctions. However, the robustness of sideband polarizations to increasing dephasing remains to be explored. Here, we investigate the dependence of high-order sideband generation in bulk gallium arsenide on dephasing by tuning temperature. We find that the intensities of the sidebands, but not their polarizations, depend strongly on temperature. Using our polarimetry method, we are able to isolate the contributions of electron-heavy hole (HH) and electron-light hole (LH) pairs to sideband intensities, and separately extract the nonequilibrium dephasing coefficients associated with the longitudinal optical (LO) phonons and acoustic (A) phonons for each species of electron-hole pair. We find that $\Gamma_{\text{HH},\text{A}} = 6.1 \pm 1.6$ $\mu$eV/K, $\Gamma_{\text{LH},\text{A}} < 1.5$ $\mu$eV/K, $\Gamma_{\text{HH},\text{LO}} = 14 \pm 3$ meV, and $\Gamma_{\text{LH},\text{LO}} = 30 \pm 3$ meV.
\end{abstract}

\maketitle

%\tableofcontents

\section{\label{intro}Introduction}
One of the chief aims of modern condensed matter physics is to study the coherent properties of quasiparticles in the form of Bloch waves in crystalline solids. In contrast with atomic and photonic systems, the dynamics of coherence in solids is complicated by interactions between the Bloch waves of interest and other quasiparticles. These interactions can lead to dephasing which destroys quantum coherence. The importance of dephasing processes in materials has led to a great deal of interest in understanding the role of dephasing in crystals in general~\cite{skinner_theory_1988, hsu_nonperturbative_1984, hsu_nonperturbative_1985, hsu_nonperturbative_1985-1} and in specific systems~\cite{gornyi_dephasing_2005,gutman_nonequilibrium_2008,bagrets_relaxation_2008,konrad_temperature_2013,sekine_dephasing_2003,gammon_phonon_1995,zhang_temperature_1995, rudin_temperature-dependent_1990-1, neder_controlled_2007, yamauchi_non-equilibrium_2009, vampa_theoretical_2014, wang_quantum_2021, du_temperature-induced_2022}. % There has been particular interest in measuring dephasing in systems that can host qubits.

A powerful optical technique that is deeply affected by dephasing is interferometry. Interferometry allows experimental access to quantities of interest such as wavefunction phases and quantum exchange statistics, which are otherwise difficult to measure~\cite{neder_interference_2007, du_subcycle_2018, schmid_tunable_2021}. Because interferometric visibility relies on coherent processes, it can be reduced or extinguished by dephasing. Despite this difficulty, interference in solids has been observed in a variety of systems. In some experiments, coherences are protected by engineering devices that host states with very small dephasing, such as chiral edge states of two-dimensional electron gases in quantum wells~\cite{neder_coherence_2007, neder_controlled_2007, neder_entanglement_2007, neder_interference_2007}. Other coherent processes are detected over time periods that are comparable to or shorter than the dephasing time. For example, in high harmonic generation (HHG) in solids, an intense laser drives charge carriers and causes the emission of harmonics with frequencies that are multiples of the driving laser frequency. The driving laser is sufficiently strong that the harmonics are generated before the carriers are dephased \cite{jin_michelson_2018, du_subcycle_2018, hohenleutner_real-time_2015, schubert_sub-cycle_2014, kim_spectral_2019}. 

Interferometric studies of HHG are exciting because HHG is a non-destructive process that can be applied to a wide variety of solid state systems such as traditional semiconductors~\cite{schubert_sub-cycle_2014}, ultra wide gap insulators~\cite{luu_extreme_2015}, and topological insulators~\cite{schmid_tunable_2021}. Depending on the materials and driving fields, different intraband trajectories, multiple interband ionization pathways, or the presence of both intraband and interband contributions can lead to quantum interferences~\cite{du_subcycle_2018, hohenleutner_real-time_2015, kim_spectral_2019, vampa_linking_2015, ndabashimiye_solid-state_2016, vampa_theoretical_2014}. However, the richness of microscopic interference sources makes it difficult to attribute a given HHG signal to any particular contribution.

High-order sideband generation (HSG) is a nonlinear optical process similar to the interband processes in HHG. HSG occurs in semiconductors when a relatively weak near-infrared (NIR) laser tuned to the bandgap creates an electron-hole pair, which is then accelerated by a strong low-frequency laser. The recombination of the electron and hole results in the emission of sideband photons. Unlike HHG, ballistic recollisions of electron-hole pairs are the only known coherent processes in HSG~\cite{zaks_experimental_2012}. The absence of other mechanisms makes the analysis of HSG relatively simple. If there are multiple bands that are degenerate or close in energy, then these bands can all contribute to the sideband signals. These different contributions interfere, influencing the polarizations of the sidebands. A newly developed technique called high-order sideband polarimetry, has been used to reconstruct the Bloch wavefunctions of holes in bulk gallium arsenide (GaAs)~\cite{costello_reconstruction_2021}, measure the effects of the dynamic phases of Bloch waves in GaAs~\cite{ohara_bloch-wave_2023}, and probe Berry curvatures in GaAs quantum wells~\cite{banks_dynamical_2017}. Based on polarimetry of high-order sidebands, methods of directly extracting the dephasing rates and information about band structures have also been proposed~\cite{wu_explicit_2023}. HSG has been demonstrated in several different materials up to room temperature~\cite{zaks_high-order_2013, borsch_super-resolution_2020, langer_lightwave-driven_2016, langer_lightwave_2018}. However, the robustness of sideband polarization in HSG to increasing dephasing has not been investigated. Specifically, in bulk GaAs and GaAs-based quantum wells, quantum interferences in sideband polarizations have been studied only at low temperature, where dephasing is minimal~\cite{zaks_experimental_2012, banks_terahertz_2013}.

Dephasing of interband excitations in GaAs has been studied previously, in both theory and experiment~\cite{hwang_lifetimes_1973, becker_femtosecond_1988, gopal_photoluminescence_2000, rudin_temperature-dependent_1990, alperovich_influence_1976, wang_transient_1993}. Experimentally, this is normally done by examining the width of optical peaks, for example, in linear absorption and photoluminescence. These experiments are complicated by the fact that GaAs has two species of holes, heavy holes (HH) and light holes (LH). These species of holes are degenerate at the band edge. This makes it difficult to experimentally distinguish the dephasing rates of their corresponding electron-hole pairs, and to our knowledge previous experiments have not been able to separately measure them.

Here, we investigate the role of dephasing in HSG in bulk GaAs by tuning the temperature from 25 K to 200 K. We perform these experiments in the low-density regime where interactions between photo-generated electron-hole pairs can be ignored, so the primary sources of temperature-dependent dephasing come from interactions with acoustic and longitudinal optical (LO) phonons~\cite{scholz_holephonon_1995}. Since high-order sideband polarimetry can distinguish between contributions to sideband polarization from different hole species, we are able to analyze temperature-dependent sideband polarization data to separately extract the temperature-dependent dephasing rates of the electron-HH and electron-LH pairs and determine the dephasing coefficents associated with the acoustic phonons and LO phonons for each species of electron-hole pair. 

\begin{figure*}
\includegraphics[width=\textwidth]{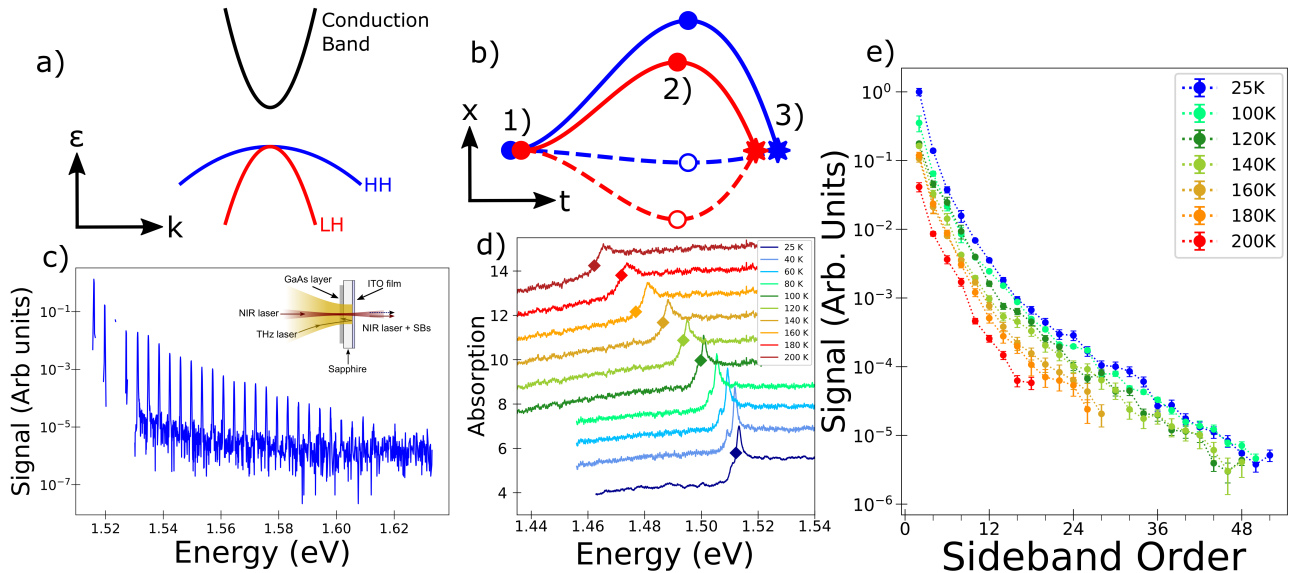}
\caption{\label{fig:one}\textbf{Temperature dependent high-order sideband generation (HSG).} \textbf{(a)} The band structure near the $\Gamma$ point ($\textbf{k}=0$) of gallium arsenide (GaAs). There is an electron (E) band, a heavy-hole (HH) band, and a light-hole (LH) band. The bandgap is shrunk by several orders of magnitude for visualization purposes. In HSG, electron-hole pairs are created at the $\Gamma$ point and accelerated to higher quasimomenta such that when annihilation occurs sideband photons are emitted at higher energies. \textbf{(b)} Real space trajectories of electrons and holes that contribute to the same order of sideband. 1) The E-HH (blue) and E-LH (red) pairs are created by the near-infrared (NIR) laser, 2) and are accelerated by a strong linearly polarized terahertz (THz) field. 3) The electrons and holes recombine, releasing a photon at a higher energy. \textbf{(c)} A sideband spectrum measured at 25 K with the NIR laser polarized parallel to the THz field. Inset: a cartoon of the experimental setup. The NIR and THz lasers are focused collinearly on an epilayer of GaAs mounted onto a sapphire substrate. The THz field is reflected by an indium tin oxide (ITO) layer which creates an optical enhancement cavity for the THz field. The NIR laser and sideband fields are transmitted through the ITO to be measured. \textbf{(d)} Temperature-dependent absorbance spectra from 25 K (blue) to 200 K (red) in dB, with diamonds representing the NIR photon energies used in HSG experiments at each temperature. For further details on this measurement, see Appendix~\ref{App:Calibration}. Each spectrum is offset by a 1 dB step for visiblity. \textbf{e)} Sideband intensities as functions of sideband order at different temperatures ranging from 25 K (blue) to 200 K (red). Experiments were performed at 25 K ($\hbar\omega_{\text{NIR}}=1.512$ eV), 100 K ($\hbar\omega_{\text{NIR}}=1.500$ eV), 120 K ($\hbar\omega_{\text{NIR}}=1.493$ eV), 140 K ($\hbar\omega_{\text{NIR}}=1.486$ eV), 160 K ($\hbar\omega_{\text{NIR}}=1.477$ eV), 180 K ($\hbar\omega_{\text{NIR}}=1.472$ eV), and 200 K ($\hbar\omega_{\text{NIR}}=1.462$ eV). The intensities are normalized to that of the 2nd order sideband at 25 K. In \textbf{(c)} and \textbf{(e)}, the NIR laser field is polarized parallel to the THz electric field, and only the components along the THz polarization are shown. See Appendix~\ref{App:Stokes} for details.}
\end{figure*}
\section{\label{hsgpol}High-order Sideband Polarimetry}

The experiments described here were performed in a bulk gallium arsenide (GaAs) epilayer. A near-infrared (NIR) laser was used to create electron-hole pairs and a strong terahertz (THz) laser was used to drive their recollisions. The THz source used in this paper was the UCSB millimeter-wave Free Electron Laser (FEL)~\cite{ramian_new_1992}, with a frequency of $449\pm 1$ GHz~\cite{valovcin_optical_2018}, a field strength of $65 \pm 3$ kV/cm at the GaAs epilayer, and a pulse length of $40$ ns. The electron-hole pairs were accelerated by the THz field in the (001) plane of GaAs. The energies of sidebands are 
\begin{equation}\label{eq:sb_en}
    \hbar \omega_{\text{SB},n} = \hbar \omega_{\text{NIR}}+n\hbar \omega_{\text{THz}},
\end{equation}
where $\hbar$ is the reduced Planck constant, $n$ is an integer called the sideband order, $\omega_{\text{SB},n}$ is the frequency of the nth-order sideband, $\omega_{\text{NIR}}$ is the frequency of the NIR laser, and $\omega_{\text{THz}}$ is the THz frequency. Because of the inversion symmetry of the (001) plane in GaAs, $n$ must be even.

Figure~\ref{fig:one} (a) shows the relevant band structure of GaAs with one electron (E) band, a heavy-hole (HH) band, and a light-hole (LH) band~\cite{luttinger_motion_1955}. Due to the degeneracy of the HH and LH bands at the band edge, the NIR laser creates both E-HH and E-LH pairs, which are then accelerated by the THz field. For a sideband of sufficiently low energy, there are classical recollision pathways associated with both the E-HH and E-LH pairs, as shown in Fig.~\ref{fig:one} (b). Depending on when the electron-hole pairs are created, they are annihilated at different times with different kinetic energies, leading to emission of different sidebands. Figure~\ref{fig:one} (c) shows a typical sideband spectrum, with sideband orders from $n=2$ to $n=52$ visible. This particular spectrum was taken at $\lambda_{\text{NIR}}=820$ nm with the NIR laser polarized parallel to the THz field. The inset of Fig. \ref{fig:one} (c) shows the experimental setup of our experiments, with the NIR and THz lasers collinearly focused on a $500$ nm epitaxially grown GaAs that was mounted onto a sapphire substrate. An indium tin oxide (ITO) film, which reflects the THz field to create an enhancement cavity and is transmissive to the NIR laser and sideband fields, was deposited onto the opposite side of the substrate. The sample is the same one as used in Ref.~\cite{ohara_bloch-wave_2023} and is prepared in the same way as described in Ref.~\cite{costello_reconstruction_2021}. On top of the ITO film, there is a silicon dioxide layer acting as an anti-reflection coating for the NIR laser and sideband fields.

Each sideband contains contributions from the two species of holes, which have different angular momenta. Since the total angular momentum must be conserved in the electron-photon interaction, this difference results in different photon helicity produced by different types of electron-hole pair. Therefore, sideband polarizations can be thought of as resulting from interferences between the different quantum paths followed by the E-LH and E-HH pairs.

The falloff of sideband intensity with increasing sideband order is controlled by dephasing. Previous work has established that HSG in bulk GaAs persists up to 170 K~\cite{zaks_high-order_2013}, but no previous study has investigated the polarizations of these sidebands at temperatures above 60K. The analysis of HSG is greatly simplified by tuning the NIR laser to the bandgap $E_{\rm g}$, i.e., $\hbar\omega_{\text{NIR}}=E_{\rm g}$~\cite{costello_reconstruction_2021}. In this way all electrons and holes are created at quasimomentum ${\bf k}=0$. However, the bandgap decreases with increasing temperature. Figure~\ref{fig:one} (d) shows the optical absorption of our sample at different temperatures. The peaks get broader with increasing temperature, which is indicative of stronger near-equilibrium dephasing at higher temperatures. 

We performed high-order sideband polarimetry at the temperatures listed in Fig.~\ref{fig:one} (d) and Fig.~\ref{fig:one}(e). In HSG experiments, the NIR laser was tuned to the temperature-dependent bandgap of GaAs at each temperature. Figure~\ref{fig:one} (e) shows the sideband intensity as a function of sideband order for all temperatures we investigated. For higher temperatures, one expects that the stronger dephasing will result in sideband intensities that are generally lower, and hence fewer sidebands will be observed. However, below 160 K, we observed similar numbers of sidebands. This is encouraging for applications of HSG in broadband frequency combs~\cite{valovcin_optical_2018}. Our results indicate that cooling down to below liquid-nitrogen temperatures is not necessary to obtain acceptable sideband signals.

\section{\label{temppol}Temperature Dependence of Sideband Polarization}
\begin{figure}
\includegraphics[width=\columnwidth]{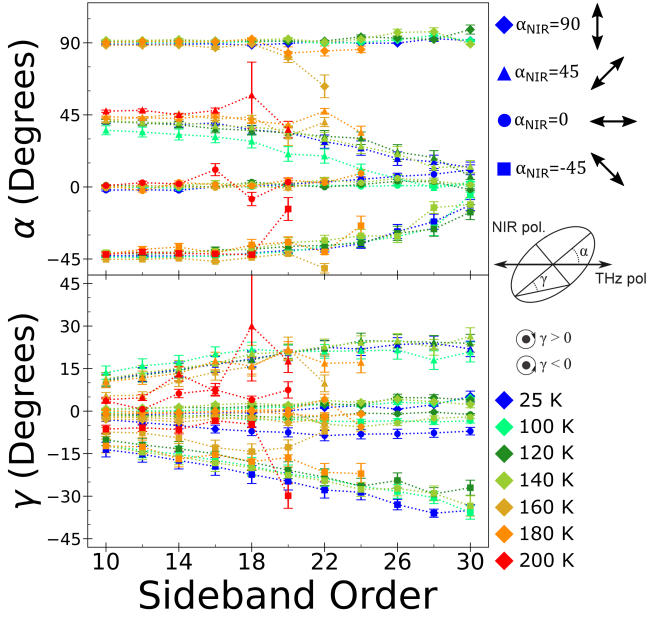}
\caption{\label{fig:two}\textbf{Temperature-dependent HSG polarimetry.} The linear orientation angle $\alpha$ (upper panel) and ellipticity angle $\gamma$ (lower panel) of sidebands are plotted as functions of sideband order for temperatures ranging from 25 K (blue) to 200 K (red). The colors match those in Fig.~\ref{fig:one} (d) and (e). The angles $\alpha$ and $\gamma$ are defined with respect to the THz field (cartoon on the right). For each temperature, data for four linear polarizations of the NIR laser ($\gamma_{\text{NIR}}=0^{\circ}$) are shown, with $\alpha_{\text{NIR}}=90^{\circ}$ (diamonds), $\alpha_{\text{NIR}}=45^{\circ}$ (triangles), $\alpha_{\text{NIR}}=0^{\circ}$ (circles), and $\alpha_{\text{NIR}}=-45^{\circ}$ (squares). Above 140 K, fewer sidebands are measured. At 200 K, no sidebands were measured for $\alpha_{\text{NIR}}=90^{\circ}$. See Appendix~\ref{App:Stokes} for details on polarization measurement.}
\end{figure}
We measure the polarizations of sidebands using the method of Stokes polarimetry (see Appendix~\ref{App:Stokes} for details). This method is sensitive to both the linear orientation angle $\alpha$ and the ellipticity angle $\gamma$ (see the polarization ellipse on the right-hand side of Fig.~\ref{fig:two}). Numerically, we constrain these angles to be in the ranges $-90^{\circ} \leq \alpha \leq 90^{\circ}$ and $-45^{\circ} \leq \gamma \leq 45^{\circ}$. A linear polarization corresponds to $\gamma=0$, and right-handed (left-handed) circularly polarized light corresponds to $\gamma=45^{\circ}$ ($\gamma=-45^{\circ}$).

Figure~\ref{fig:two} shows the polarimetry results for all temperatures we investigated. At each temperature, four polarimetry scans were performed for four different linear polarizations of the NIR laser. The four linear orientation angles of the NIR laser were $\alpha_{\text{NIR}}= 90^{\circ}$ (diamonds), $\alpha_{\text{NIR}}= 45^{\circ}$ (triangles), $\alpha_{\text{NIR}}= 0^{\circ}$ (circles), and $\alpha_{\text{NIR}}= -45^{\circ}$ (squares). The upper and lower panels show the $\alpha$ and $\gamma$ angles for the sidebands. The sideband polarizations exhibit remarkable robustness to temperature, with small variations up to 160 K. Above 160 K, reduced signal-to-noise ratio leads to more deviation in higher-order sidebands, but even at 200 K the polarizations of the lower-order sidebands are very similar to those at lower temperatures. 

These results reinforce our understanding that the sideband polarizations are set by the interference of Bloch waves. In our HSG experments, this interference arises from different dynamic phases accumulated by the E-HH (E-LH) pair from the creation time $t_{0}$ to annihilation time $t_{f}$,
\begin{equation}\label{eq:dyn_phaes}
    A_{\text{HH}(\text{LH})}(t_{f}, t_{0}) = -\frac{1}{\hbar} \int_{t_0}^{t_f} dt' E_{\text{HH}(\text{LH})}(t').
\end{equation}
Here, $t_{0}$ and $t_{f}$ depend on the sideband order as well as the species of electron-hole pairs. The relative energy of the E-HH and E-LH pairs, $E_{\rm HH}$ and $E_{\rm LH}$, contain the bandgap $E_{\rm g}$, which is the only temperature-dependent factor in the dynamic phases. Since the NIR laser is tuned to the bandgap at each temperature, this temperature dependence is not relevant for the sideband polarization~\cite{costello_reconstruction_2021}. The dephasing, however, does depend strongly on temperature. This suggests that the electron-hole coherences are damped faster at higher temperature by the stronger dephasing, resulting in the lower sideband intensities, as shown in Fig.~\ref{fig:one} (e). In addition, the electron-hole pairs that survive until annihilation and sideband emission experience the same coherent acceleration, and therefore the information contained in the sideband polarization is nearly independent of temperature.

\section{\label{deph}Extracting Dephasing Rates from Sideband Polarimetry}

\begin{figure*}
\includegraphics[width=\textwidth]{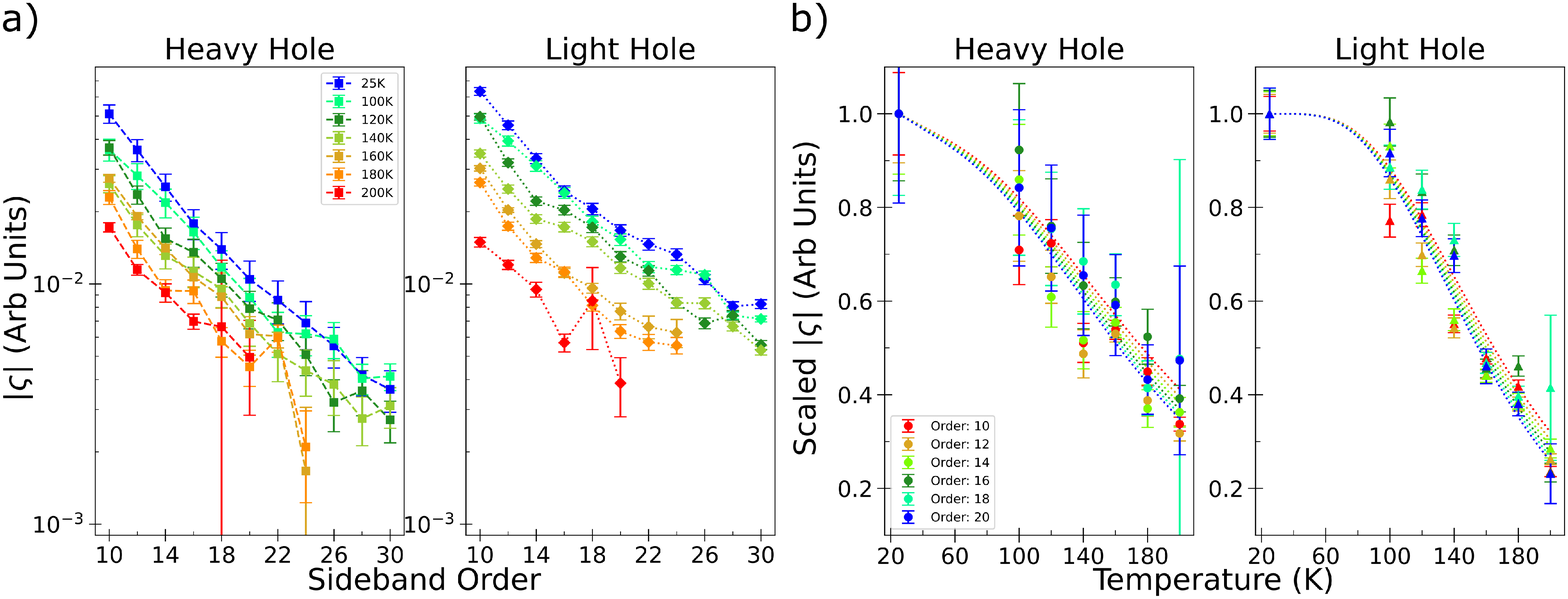}
\caption{\label{fig:three}\textbf{Temperature-dependent falloffs of the E-HH and E-LH contributions to sideband intensity.} \textbf{(a)} The absolute values of the propagators for the E-HH and E-LH pairs, $\varsigma_{\text{HH}}$ (left, squares) and $\varsigma_{\text{LH}}$ (right, diamonds) for different temperatures running from 25 K (blue) to 200 K (red) as functions of sideband order. The colors match those in Fig.~\ref{fig:two}. \textbf{(b)} The absolute values of $\varsigma_{\text{HH}}$ (left, circles) and $\varsigma_{\text{LH}}$ (right, triangles) of different sideband orders as functions of temperature. Each plot is normalized to the 25 K data to isolate the dependence of the falloff for each sideband on temperature. The dotted lines in each frame show the results from fitting all data for each hole species to Eq.~\ref{eq:dephase_prop}. For each fit, there are two free parameters, $\Gamma_{j,\text{A}}$ and $\Gamma_{j,\text{LO}}$, where $j$ labels the hole species. Different sideband orders are associated with different acceleration time $\tau_{j}(n)$ for the recollision processes, so the dotted lines for different sideband orders are slightly different despite the dephasing function $\Gamma_{j}(T)$ being the same. }
\end{figure*}

Although the polarization of sidebands does not depend strongly on temperature, the intensities of these sidebands can carry information on the temperature dependence of dephasing. By using the results of Ref.~\cite{costello_reconstruction_2021}, the temperature dependence of the dephasing rates of the E-LH and E-HH pairs can be extracted separately. The electric field of the $n$th-order sideband can be written as 
\begin{equation}\label{eq:elec_pol}
    \mathbb{P}_{n} \propto 
    \sum_{s} 
    \begin{pmatrix}
    {\bf D}_{\text{E-HH},s} \\
    {\bf D}_{\text{E-LH},s}
    \end{pmatrix}^{\dagger}
    \begin{pmatrix}
    \varsigma_{\text{HH}, n} & 0 \\
    0 & \varsigma_{\text{LH}, n} 
    \end{pmatrix}
    \begin{pmatrix}
    {\bf D}_{\text{E-HH},s} \\
    {\bf D}_{\text{E-LH},s}
    \end{pmatrix} \cdot {\bf F}_{\text{NIR}}
\end{equation}
where $s$ labels the two-fold degeneracy in the electron-hole states, ${\bf D}_{\text{E-HH (E-LH)},s}$ is the electric dipole vector associated with the E-HH (E-LH) pair labeled by $s$, ${\bf F}_{\text{NIR}}$ is the electric field of the NIR laser, and $\varsigma_{\text{HH}(\text{LH}), n}$ is the propagator of the E-HH (E-LH) pair. The dynamics of electron-hole recollisions is captured by the propagator $\varsigma_{j, n}$ ($j={\rm HH,LH}$), which assumes the form,\begin{equation}\label{eq:sigma_int}
    \varsigma_{j, n} \propto \int_{0}^{2\pi/\omega_{\rm THz}}dt e^{in\omega_{\rm THz} t} \int_{-\infty}^{t}dt' e^{iA_{j}(t,t')-(t-t')\Gamma_{j}(T)/\hbar},
\end{equation}
where $A_{j}(t,t')$ is the dynamic phase defined by Eq.~\ref{eq:dyn_phaes}, and $\Gamma_{j}(T)/\hbar$ is the dephasing rate written as a function of temperature $T$. The first integral is a Fourier transform to isolate the electron-hole pairs which have gained kinetic energy $n\hbar\omega_{\rm THz}$ upon recombination and therefore contribute to the $n$th order sideband. The second integral sums over the contributions from electron-hole pairs created at different times $t'$. 
In this expression, only the recollision pathways associated with electron-hole pairs created at zero quasimomentum are included. Because the NIR laser in our HSG experiment is relatively weak, the population of electrons and holes is not high enough for carrier-carrier scattering effects to be an important source of dephasing. We ensure that this is the case by confirming that the sideband intensities are linearly proportional to the NIR laser power (we used 100 mW in these experiments). We model the dephasing function $\Gamma_{j}(T)$ ($j={\rm HH,LH}$) as
\begin{equation}\label{eq:dephase_def}
    \Gamma_{j}(T)=\Gamma_{0}+\Gamma_{j,\text{A}}T+\Gamma_{j,\text{LO}}n_{\text{LO}}(T),
\end{equation}
which includes a temperature-independent dephasing constant $\Gamma_{0}$, and contributions from the long-wavelength acoustic phonons and the longitudinal optical (LO) phonons proportional to the phonon occupation numbers. We assume that the occupation number for the long-wavelength acoustic phonons is proportional to temperature, while
the occupation number for the long-wavelength LO phonons is taken as Bose-Eistein distribution, $n_{\rm LO}(T)=1/[\exp(\hbar\omega_{\rm LO}/k_{\rm B}T)-1]$, where $\hbar\omega_{\rm LO}=36.6$ meV is the energy of the LO phonons~\cite{banks_terahertz_2013} and $k_{\rm B}$ is the Boltzmann constant. We refer to $\Gamma_{j,\text{A}}$ and $\Gamma_{j,\text{LO}}$ as dephasing coefficients since both need to be multiplied by additional factors to give the dephasing rate. 

Figure~\ref{fig:three} (a) shows the absolute values of the propagators, $|\varsigma_{\text{HH},n}|$ and $|\varsigma_{\text{LH},n}|$, as functions of sideband order. As expected, the propagators have larger amplitudes at lower temperatures. Intriguingly, the falloffs of the propagators for the E-HH and E-LH pairs are slightly different. This is due to slight differences in dephasing rates, creation times, and annihilation times for the E-HH and E-LH pairs.

We assume here that the dephasing rate $\Gamma_{j}/\hbar$ is independent of quasimomentum, and can be understood as the mean dephasing rate experienced by the electron-hole pairs during acceleration from the THz field. This assumption implies that the dephasing rate is constant in time for a given electron-hole pair, which allows us to isolate the dephasing dependence in $|\varsigma_{j,n}|$ as 
\begin{equation}\label{eq:dephase_prop}
    |\varsigma_{j,n}| \propto e^{ -\Gamma_{j}(T)\tau_{j}(n)/\hbar },
\end{equation}
where $\tau_{j}$ is the time between the creation and annihilation of the electron-hole pair labeled by $j$ in the shortest classical recollision pathway (referred to as the acceleration time of the electron-hole pairs). We have assumed that the dephasing investigated here is sufficiently strong such that, for each species of the electron-hole pair, there is one shortest recollision pathway that dominantly determines the polarization of a sideband. To simplify the analysis, we consider here classical recollisions, in which the electrons and holes are created and annihilated at the same position. In our HSG experiment, the average kinetic energy gain of the electron-hole pairs in a THz cycle, namely, the ponderomotive energy $U_{{\rm p},j}={e^{2}F^{2}_{\rm THz}}/{ 4\mu_j \omega_{\rm THz}^{2} }$ ($j={\rm HH,LH}$), is much larger than the sideband offset energies and the dephasing function $\Gamma_{j}(T)$. Here, $e$ is the elementary charge, $F_{\rm THz}$ is the THz field strength, and $\mu_{\rm HH}=[m_c^{-1}+\gamma_1-2\gamma_2|{\bf n}|]^{-1}$ ($\mu_{\rm LH}=[m_c^{-1}+\gamma_1+2\gamma_2|{\bf n}|]^{-1}$) is the reduced mass of the E-HH (E-LH) pairs along the linear polarization of the THz field. The parameter $m_c=0.067m_0$ is the effective mass of the conduction band ($m_0$ is the electron rest mass), $\gamma_{1}=6.98$, $\gamma_{2}=2.06$ are two Luttinger parameters~\cite{vurgaftman_band_2001}, and the vector ${\bf n} = \left( ({\sqrt{3}}/2)\sin 2\theta, - ({ \sqrt{3}\gamma_{3} }/{ 2\gamma_{2}) }\cos 2\theta, -{1}/{2} \right)$ is defined by the Luttinger parameters $\gamma_{2}$ and $\gamma_{3}=2.93$~\cite{vurgaftman_band_2001}, together with the angle between the linear polarization of the THz field and the [110] crystal direction of GaAs, $\theta=67^{\circ}$. As discussed in Ref.~\cite{wu_explicit_2023}, in this case, the THz field can be approximated as linear in time during the recollision processes of the electron-hole pairs. With this approximation, the acceleration time associated with the $n$th-order sideband can be explicitly calculated as~\cite{wu_explicit_2023,ohara_bloch-wave_2023} 
\begin{align}\label{lit_times}
        \tau_{j}(n) = \frac{\sqrt{3}}{\omega_{\rm THz}}\left( \frac{2n\hbar\omega_{\rm THz}}{U_{{\rm p},j}} \right)^{1/4}, 
\end{align}
The acceleration times $\tau_{j}$ are different for the E-HH and E-LH pairs because of their different reduced masses, leading to different trajectories that cause quantum interferences. 
%In Eq.~\ref{eq:dephase_prop} it is clear that the dephasing  depends only on temperature and the acceleration time depends only on sideband order. 
%The acceleration time can be calculated from experimental parameters, see Appendix \ref{App:LiT}. Here the THz field strength is strong enough that we can simplify our analysis by using the THz linear-in-time (LIT) approximation, where the creation and annihilation times are calculated assuming that the THz field is linear in time \cite{wu_explicit_2023, ohara_bloch-wave_2023}. 

Figure~\ref{fig:three} (b) shows $|\varsigma_{j,n}|$ as a function of temperature for sideband orders from $n=10$ to $n=20$. For each sideband order, the values of $|\varsigma_{j,n}|$ are normalized to its value at 25 K. This normalization allows all of the data to be fit by exponential functions (Eq.~\ref{eq:dephase_prop}) with only 2 free parameters for each hole species, $\Gamma_{j,\text{A}}$ and $\Gamma_{j, \text{LO}}$. The dotted lines in Fig.~\ref{fig:three} (b) show the fitting results. There are different lines for different sideband orders because of the difference in the acceleration time $\tau_{j}(n)$.

\begin{figure}
\includegraphics[width=\columnwidth]{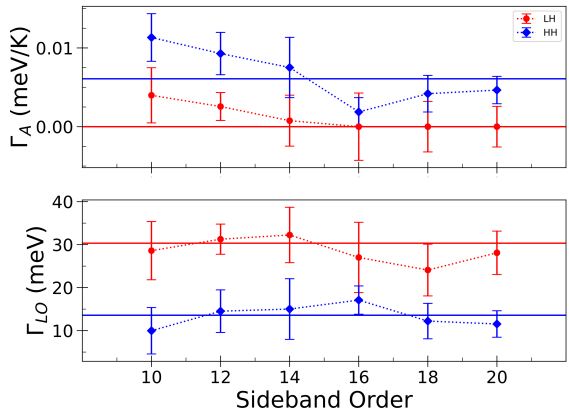}
\caption{\label{fig:four}\textbf{Extracted coefficients for the temperature dependent part of dephasing rates.} The coefficients $\Gamma_{j,\text{A}}$ and $\Gamma_{j,\text{LO}}$ in the dephasing function (Eq.~\ref{eq:dephase_def}) are extracted by fitting to scaled $|\varsigma_{\text{HH}}|$ and $|\varsigma_{\text{LH}}|$ shown in Fig.~\ref{fig:three} (b) with Eq.~\ref{eq:dephase_prop}. The red (blue) solid horizontal lines represent the results from fitting to the data for all sideband orders with a dephasing function $\Gamma_{\text{LH}}(T)$($\Gamma_{\text{HH}}(T)$) that is assumed to be independent of the sideband order. The numerical values are $\Gamma_{\text{HH},\text{A}} = 6.1 \pm 1.6$ $\mu$eV/K, $\Gamma_{\text{LH},\text{A}} < 1.5$ $\mu$eV/K, $\Gamma_{\text{HH},\text{LO}} = 14 \pm 3$ meV, and $\Gamma_{\text{LH},\text{LO}} = 30 \pm 3$ meV, which produce the dotted lines in Fig.~\ref{fig:three} (b). These error bars are the standard deviations by taking the square roots of the variance of the fits.}
\end{figure}

Figure~\ref{fig:four} shows the parameters $\Gamma_{j,\text{A}}$ and $\Gamma_{j, \text{LO}}$ that are returned by the fitting procedure. The diamonds (dots) refer to fits for each sideband order individually for the E-HH (E-LH) pairs, and the solid lines show the results from fitting to the data for all sideband orders simultaneously. The results of the fit are $\Gamma_{\text{HH},\text{A}} = 6.1 \pm 1.6$ $\mu$eV/K, $\Gamma_{\text{LH},\text{A}} < 1.5$ $\mu$eV/K, $\Gamma_{\text{HH},\text{LO}} = 14 \pm 3$ meV, and $\Gamma_{\text{LH},\text{LO}} = 30 \pm 3$ meV. The errors reported are the square roots of the variances of the fit. The fits are likely less reliable for the dephasing coefficents associated with the acoustic phonons since the contribution from the LO phonon dominates for most of the sampled temperature range. This result agrees with previous results that the LO phonon is the dominant source of dephasing in bulk GaAs above 80 K~\cite{gopal_photoluminescence_2000, banks_terahertz_2013}. 

It is important to note that the coefficients $\Gamma_{j,\text{LO}}$ must be multiplied by the LO phonon occupation number to give their contribution to $\Gamma_{j}(T)$ in Eq.~\ref{eq:dephase_def}. Thus, for example, at 100 K, where the phonon occupation number is 0.015, the LO-phonon contributions to $\Gamma_{j}(T)$ are $\Gamma_{\rm HH,LO}n_{\rm LO}= 0.21$ meV and $\Gamma_{\rm HH,LO}n_{\rm LO}= 0.45$ meV. For comparison, $\Gamma_{j}(T=35 {\rm K})$, which is close to $\Gamma_0$ because the LO phonon occupation is near 0, has been estimated to be $4.8\hbar\omega_{\rm THz}=9$ meV~\cite{ohara_bloch-wave_2023}.

%For comparison, other studies have estimated the total dephasing at 35 K to be roughly 9 meV with the same experimental setup~\cite{ohara_bloch-wave_2023}. At this temperature, the effects of electron-phonon interaction should be negligible due to small phonon occupation numbers and the total dephasing result can be used to estimate the value of $\Gamma_{0}$ in Eq.~\ref{eq:dephase_def}. At 200 K $n_{\text{LO}} \approx 0.1$, leading to total dephasings of roughly $\Gamma_{\text{HH}}= 11.6 \text{meV}$ and $\Gamma_{\text{LH}}= 12.3 \text{meV}$.

\section{\label{disc}Discussion}
Our results are not the first to measure dephasing in GaAs, but our method based on sideband polarimetry represents a new way to measure dephasing that differs from previous experiments. The ability of our method to distinguish the contributions from different Bloch waves is a key advantage. Although theory has long predicted that the scattering and dephasing rates should differ for different hole species~\cite{scholz_holephonon_1995, rudin_temperature-dependent_1990}, this is to our knowledge the first experiment that differentiates the dephasing coefficients of these holes near the bandgap. Previous investigations of dephasing in GaAs have used the linewidth of photoluminescence or absorption spectra, and have been complicated by the degeneracy of the LH and HH band at the band edge~\cite{rudin_temperature-dependent_1990, gopal_photoluminescence_2000}. These experiments found $\Gamma_{\text{A}} = 13 \pm 3$ $\mu$eV/K and $\Gamma_{\text{LO}} = 30.4 \pm 4$ meV, but they were unable to differentiate HH from LH contributions, and had little confidence in the value of $\Gamma_{\text{A}}$~\cite{gopal_photoluminescence_2000}. These methods have particular difficulty at high temperatures, where strain-induced splittings of the HH and LH bands are washed out by broadened peaks. Four-wave mixing has also been used to probe dephasing in GaAs~\cite{wang_transient_1993}, but is similarly only able to examine low-temperature physics. It is interesting that these near-equilibrium measurements seem to agree with our measurement of $\Gamma_{\text{LH},\text{LO}}$. This is perhaps due to the fact that the LH peak is broader due to higher dephasing, and could dominate these linewidth measurements.

Theoretical explorations of $\Gamma_{\text{LO}}$ in GaAs have found that dephasing of excitons depends heavily on the effective mass of the hole, with lower hole masses corresponding to higher dephasing rates~\cite{rudin_temperature-dependent_1990}. This study found $\Gamma_{\text{HH},\text{LO}}= 21$ meV for excitons in near equilibrium. The difference from our results could come from the fact that the dephasing experienced by the electron-hole pairs in our HSG experiment are in continuum states far from equilibrium. Indeed, the methods in Ref.~\cite{rudin_temperature-dependent_1990} provide insights only into the near-equilibrium dephasing. Recently, dephasing in systems far from equilibrium in the context of HHG has been studied theoretically~\cite{wang_quantum_2021}, including effects of phonons~\cite{du_temperature-induced_2022}. However, to our knowledge, species-specific nonequilibrium dephasing rates in semiconductors have not been previously measured. Our results on GaAs pave the way to measuring the nonequilibrium dephasing rates of a large class of semiconducting materials.

A further advantage of our method is the possibility of measuring the dependence of dephasing on the direction of electron-hole acceleration. By rotating the THz polarization with respect to the crystal, the electron-hole pairs will be driven along different directions. By studying this angular dependence of sideband polarization, it is possible to probe dephasing in different crystal directions. This sensitivity is not possible in linear optical methods such as linear absorption. Future experiments could be particularly useful in probing the directional dephasing of systems that have significant anisotropy. In GaAs, changing the direction of electron-hole acceleration could probe the dependence of dephasing on effective masses of holes, since the effective masses of both the HH and LH depend on crystal orientation.

\section{\label{conc}Conclusion}
As application of the quantum effects in materials advances, experimental knowledge of the coherent properties of Bloch wavefunctions becomes more critical. Polarimetry of high-order sidebands has been shown as an exciting interferometry method that provides insight into the phases of Bloch wavefunctions and allows extraction of key quantities in electronic structures~\cite{costello_reconstruction_2021, ohara_bloch-wave_2023, wu_explicit_2023, langer_lightwave_2018, langer_lightwave-driven_2016, borsch_super-resolution_2020}. The fact that these polarization states are robust over a wide range of temperatures is encouraging both to the viability of further low cost HSG experiments in GaAs and to the extension of HSG to other materials which have greater dephasing rates. Furthermore, in devices that take advantage of coherent quantum properties, knowledge of dephasing is critical. We have demonstrated that temperature-dependent high-order sideband polarimetry can be used to differentiate the temperature-dependent dephasing of different species of electron-hole pairs, even if they are very close in energy. Future experiments can complete this analysis by examining the temperature-independent contribution to dephasing, and by extending this method to other materials that host more exotic wavefunctions.

\begin{acknowledgments}
We acknowledge N. Agladze for assistance in maintaining and operating the UCSB millimeter-wave free electron laser. This work was funded by NSF DMR 2004995. J.B. Costello and M.S. Sherwin conceptualized this work. Q. Wu produced the theory. J.B. Costello, S.D. O'hara, and M. Jang performed the experiments. J.B. Costello performed data analysis. L.N. Pfeiffer and K.W. West grew the GaAs crystal. M.S. Sherwin acquired funding and administered the project.
\end{acknowledgments}

\appendix

%\section{\label{App:Sample} Sample Details}
%The sample used in these experiments follow the sample preparation in the methods section of \cite{costello_reconstruction_2021}, and is the same one used in \cite{ohara_bloch-wave_2023}. The final sample is 500 nm of epitaxially grown GaAs Van der Waals bonded to a sapphire substrate. The reverse side of the substrate has layers of indium tin oxide (ITO) and SiO2. The ITO reflects the THz to create an enhancement cavity but transmits the NIR and sidebands. The SiO2 acts as an anti-reflection coating for the NIR and sidebands.

\section{\label{App:Stokes} Stokes Polarimetry}
To measure sideband polarizations, we used a method called Stokes polarimetry, following the same procedure as described in Ref.~\cite{costello_reconstruction_2021}. Stokes polarimetry was performed by measuring the intensity of a light wave after it propagated through a rotating quarter-wave plate and a linear polarizer as a function of the rotation angle of the quarter-wave plate. In our experiments, the transmission axis of the linear polarizer was parallel to the linear polarization of the THz field defined as horizontal. The intensity $I$ of the outgoing light can be written as 
\begin{align}\label{eq:stokes_pol}
I(\varphi) = &\frac{S_{0}}{2}+\frac{S_{1}}{4}[1+\cos(4\varphi)]\notag\\
&+\frac{S_{2}}{4}\sin(4\varphi)-\frac{S_{3}}{2}\sin(2\varphi),
\end{align}
where $\varphi$ is the angle between the fast axis of the rotating quarter-wave plate and the horizontal, and $S_{0}$, $S_{1}$, $S_{2}$, and $S_{3}$ are the Stokes parameters, which define the polarization state of the light. We sampled 16 different rotation angles of the quarter-wave plate to extract the Stokes parameters of the sidebands. The Stokes parameters are related to the linear orientation angle $\alpha$ and ellipticity angle $\gamma$ through the following equations,
\begin{equation}\label{eq:alpha_def}
    \tan(2\alpha) = \frac{S_{2}}{S_{1}},
\end{equation}
\begin{equation}
    \tan(2\gamma) = \frac{S_{3}}{\sqrt{S_{1}^{2}+S_{2}^{2}}}.
\end{equation}
Because our experiments always use the Stokes polarimeter, the full intensities of the sidebands described by the Stokes parameter $S_0$ were not directly recorded. The spectra displayed in Fig.~\ref{fig:one} (e) are actually the intensities $I(\varphi)$ of the sidebands at $\varphi=0^{\circ}$ for a horizontal polarization of the NIR laser ($\alpha_{\text{NIR}}=0^{\circ}$). They should be understood as the intensities of the horizontal components of the sidebands. The intensity $I(\varphi)$ of a sideband tends to be the strongest when the NIR laser is horizontally polarized, because in this case the sidebands tend to be close to horizontally polarized and have maximum transmission through the linear polarizer in the Stokes polarimeter. Similarly, for a vertical polarization of the NIR laser ($\alpha_{\text{NIR}}=90^{\circ}$), the sidebands tend to have minimal transmission through the linear polarizer, which lowers the signal-to-noise ratio. For this reason, no sidebands were detected for the vertical polarization of the NIR laser ($\alpha_{\text{NIR}}=90^{\circ}$) at 200 K.

\section{\label{App:Calibration} Wavelength Calibration}
The sideband intensity data presented here were recorded with two different detectors, a Hamamatsu R7400U-20 photomultiplier tube (PMT) and an Andor NewtonEM  electron-multiplying charge coupled device (EMCCD). The PMT provides a time-resolved, single-pixel measurement of a narrow frequency range, while the EMCCD provides a simultaneous time-integrated measurement in many pixels, enabling an entire HSG spectrum to be recorded for each NIR laser pulse. The NIR laser was always much stronger than the sidebands, so the EMCCD tends to be saturated when we tried to measure the lower-order sidebands, which were very close to the NIR laser line. For this reason, we use the PMT to measure the lower-order sidebands and the EMCCD to measure higher-order sidebands. The Stokes polarimeter was only used with the EMCCD, so the polarizations for the low-order sidebands ($n<10$) were not measured in this work. See Ref.~\cite{costello_reconstruction_2021} for more details.

The quantum efficiencies of both the PMT and EMCCD vary with wavelength over the range used in this paper. Therefore, to compare HSG spectra at different temperatures, one needs to calibrate the quantum efficiencies of these two detectors to account for this wavelength dependence. The wavelengths and photon energies of the NIR laser we used are listed in Table~\ref{tab:wls},
\begin{table}[b]
\caption{\label{tab:wls}
The wavelengths and photon energies of the NIR laser used in the HSG experiments at each temperature.
}
\begin{ruledtabular}
\begin{tabular}{ccc}
Temperature (K)&
Wavelength (nm)&
Photon energy (eV)\\
\colrule
25 & 820.0 & 1.512\\
100 & 826.8 & 1.500\\
120 & 830.2 & 1.493\\
140 & 834.2 & 1.486\\
160 & 839.6 & 1.477\\
180 & 842.5 & 1.472\\
200 & 848.0 & 1.462\\
\end{tabular}
\end{ruledtabular}
\end{table}
The wavelength of the NIR laser was chosen to be slightly redshifted from the absorbance peaks in Fig.~\ref{fig:one} (d) for a given temperature in order to ensure that the electron-hole pairs were created at the band edge. The absorbance spectra shown in Fig.~\ref{fig:one} (d) were determined by measuring a broad spectrum of a white light source with and without the GaAs sample in the transmission path. The absorbance is calculated (in dB) as 
\begin{equation}\label{absorption_eq}
    \text{Abs} = -10~\text{log}_{10}(I_{S}/I_{0})
\end{equation}
where $I_{S}$ ($I_{0}$) is the measured intensity with (without) the GaAs sample inserted into the optics.

Sidebands up to order 30 were used in polarization measurements, so the wavelengths measured in this work (including the sidebands and NIR laser) range from 790 nm to 848 nm. To calibrate both detectors, we used a halogen white light source and a small tunable monochromator to select a wavelength range in five nm increments from 778 nm to 848 nm. The monochromator produced spectra with a wavelength range of roughly 30 nm. The power of the outgoing light from the monochromator was measured by using a frequency-independent FieldMaster power meter from Coherent lasers and compared with the signal of the PMT or EMCCD to determine the detector sensitivity at the selected wavelength. Since we used the same power of the NIR laser in all sideband measurements, knowing the detector sensitivities allows for direct comparison of sideband intensities measured at different temperatures. 

The quantum efficiencies of both the PMT and EMCCD decrease with increasing wavelength, which leads to lower sensitivites at higher temperatures. This explains why our noise floor is higher for higher temperatures in Fig.~\ref{fig:one} (e).

\section{\label{App:DatAnal} Data Analysis}
 We performed high-order sideband polarimetry for four different linear polarization of the NIR laser with $\alpha_{\text{NIR}}= 90^{\circ}$, $\alpha_{\text{NIR}}= 45^{\circ}$, $\alpha_{\text{NIR}}= 0^{\circ}$, and $\alpha_{\text{NIR}}= -45^{\circ}$, respectively. These measurements allow us to determine each of the dynamical Jones Matrices up to a constant factor~\cite{banks_dynamical_2017}. The dynamical Jones matrices link the electric fields of the sidebands, to the electric field of the NIR laser through the following equation,
\begin{equation}\label{Jones_mat}
    \begin{pmatrix}
    E_{+,n} \\
    E_{-,n} 
    \end{pmatrix} =
    \begin{pmatrix}
    T_{++,n} & T_{+-,n} \\
    T_{-+,n} & T_{--,n}
    \end{pmatrix}
    \begin{pmatrix}
        E_{+,\rm NIR} \\
        E_{-,\rm NIR}
    \end{pmatrix},
\end{equation}
where $E_{\pm,n}$ and $E_{\pm,\rm NIR}$ are the components with helicities $\pm1$ for the $n$th-order sideband and the NIR laser, respectively, and $T_{\pm\pm,n}$ are the dynamical Jones matrix elements. Following Ref.~\cite{costello_reconstruction_2021}, from Eq.~\ref{eq:elec_pol}, the dynamical Jones matrix elements can be written as
\begin{align}
T_{++,n}=T_{--,n}
&=
\frac{2}{3}(\varsigma_{{\rm HH},n}+\varsigma_{{\rm LH},n})
+
\frac{n_z}{3}(\varsigma_{{\rm HH},n}-\varsigma_{{\rm LH},n}),\\
T_{+-,n}&=\frac{n_x+in_y}{\sqrt{3}}(\varsigma_{{\rm HH},n}-\varsigma_{{\rm LH},n}),\\
T_{-+,n}&=\frac{n_x-in_y}{\sqrt{3}}(\varsigma_{{\rm HH},n}-\varsigma_{{\rm LH},n}).
\end{align}
Here, $(n_x,n_y,n_z)$ is a unit vector of the vector
\begin{equation}
    {\bf n} = \left( \frac{\sqrt{3}}{2}\sin 2\theta, - \frac{ \sqrt{3}\gamma_{3} }{ 2\gamma_{2} }\cos 2\theta, -\frac{1}{2} \right),
\end{equation}
which is determined by the Luttinger parameters $\gamma_{3}=2.93$ and $\gamma_{2}=2.06$~\cite{vurgaftman_band_2001}, and the angle between the linear polarization of the THz field and the [110] crystal direction of GaAs, $\theta=67^{\circ}$. Given the dynamical Jones matrix elements, we can then calculate the propagators $\varsigma_{{\rm HH},n}$ and $\varsigma_{{\rm LH},n}$ as
\begin{align}\label{app_varsigma_hh}
        &\varsigma_{\text{LH},n} = \frac{3}{4}\left(  T_{++,n}-\frac{2+n_z}{ \sqrt{3}  (n_{x}+in_{y}) }T_{+-,n} \right)\\
        \label{app_varsigma_lh}
        &\varsigma_{\text{HH},n} = \frac{3}{4}\left(  T_{++,n}+\frac{2-n_z}{ \sqrt{3}
        (n_{x}+in_{y}) }T_{+-,n} \right).
\end{align}

To establish confidence intervals in our polarization measurements, we measured the intensity $I(\varphi)$ of a given sideband 4 times for each rotation angle $\varphi$ of the quarter-wave plate in the Stokes polarimetry. The standard deviation calculated from these intensities gives an uncertainty for each sideband intensity $I(\varphi)$. The uncertainties in the intensity $I(\varphi)$ propagated throughout the analysis to give the uncertainties in the angles $\alpha$ and $\gamma$.
To determine standard deviations of the dynamical Jones matrix elements, we use a Monte Carlo method to randomly sample the angles $\alpha$ and $\gamma$ 10,000 times with their variances set by the Stokes polarimetry~\cite{costello_reconstruction_2021}. These deviations then propagate through Eqs.~\ref{app_varsigma_hh} and~\ref{app_varsigma_lh} to determine the uncertainties in the propagators, $\varsigma_{\text{LH}}$ and $\varsigma_{\text{HH}}$.

\bibliographystyle{apsrev4-2}
\bibliography{references}% Produces the bibliography via BibTeX.

\end{document}